\documentclass[twocolumn,showpacs,preprintnumbers,amsmath,amssymb,prl,epsf]{revtex4}
\usepackage{graphicx}
\begin{document}
\title{Dispersion spreading of biphotons in optical fibres
and two-photon interference}
\author{G. Brida$^1$, M.~V.~Chekhova$^2$,  M. Genovese$^1$, M. Gramegna$^1$ and L.~A.~Krivitsky$^{1,2}$}
\affiliation{$^1$ Istituto Elettrotecnico Nazionale Galileo
Ferraris, Strada delle Cacce 91, 10135 Torino, Italy }

\affiliation{$^2$Department of Physics, M.V.Lomonosov Moscow State
University,  Leninskie Gory, 119992 Moscow, Russia} \vskip 24pt
\begin{abstract}
\begin{center}\parbox{14.5cm}
{We present the first observation of two-photon polarization
interference structure in the second-order Glauber's correlation
function of two-photon light generated via type-II spontaneous
parametric down-conversion. In order to obtain this result,
two-photon light is transmitted through an optical fibre and the
coincidence distribution is analyzed by means of the START-STOP
method. Beyond the experimental demonstration of an interesting
effect in quantum optics, these results also have considerable
relevance for quantum communications.}
\end{center}
\end{abstract}
 \maketitle \narrowtext
\vspace{-10mm}

In many quantum communication, quantum computation and quantum
metrology ~\cite{Gisin} experiments, entangled states of light are
transmitted through optical fibres. For example, quantum key
distribution through fibres has already been demonstrated up to
100 km ~\cite{Mo}. Because of these applications, a clear
understanding of the effect of  fibre propagation on the
properties of entangled states  is highly demanded. This
understanding would be also very useful for the studies on
the foundations of quantum mechanics~\cite{Genovese}.

In particular, it is known that an entangled two-photon state, a
biphoton, spreads in time when propagating through a medium with
group velocity dispersion. More specifically, the second-order
intensity correlation function of two-photon light gets broadened
and at a sufficiently large distance, it takes the shape of the
spectrum~\cite{Chekhova},~\cite{Valencia}. It turns out that for
biphotons propagating through a dispersive medium, the shape of
the correlation function can manifest the effects of two-photon
interference.

In a typical experiment on observing two-photon interference for
type-II spontaneous parametric down-conversion (SPDC) (Fig.1),
pairs of orthogonally polarized signal and idler photons generated
in the collinear frequency-degenerate regime are sent to a
non-polarizing beamsplitter (NPBS) followed, in both arms, by
polarizers (P1, P2) and photon counting detectors (D1, D2).
 When one of the polarizers is rotated, the
counting rate of the corresponding detector does not vary, but the
coincidence counting rate varies with almost 100\% visibility
(polarisation fringes)~\cite{Shih}. Another way to observe
two-photon interference is to vary the phase between signal and
idler photons (space-time fringes)~\cite{Shih1}. Such an
experiment can also be interpreted as preparation of entangled
two-photon Bell states in the two spatial modes after the
beamsplitter~\cite{Pittman}.

In addition to space-time and polarization fringes, two-photon
interference can manifest itself in the modulation of the
coincidence counting rate due to the displacement of one or both
detectors in the near-field or far-field zone~\cite{Burlakov} or
variation of the frequency registered by one or both
detectors~\cite{Viciani}. However, to the best of our knowledge,
there have been no experiments on observing two-photon
interference in the shape of the second-order intensity
correlation function for SPDC, although a related experiment has
been recently performed for a narrow-band two-photon source based
on an OPO operating below threshold \cite{Ou}. The main reason is
that normally, the width of the correlation function for SPDC
radiation is on the order of hundreds of femtoseconds, and it
cannot be resolved using the existing equipment. However, after a
fibre of length about several hundred meters, the width of the
correlation function grows to several nanoseconds, which enables
the interference to be observed.

In a setup shown in Fig.1, the state vector of the biphoton field
at the output of the beamsplitter is

\begin{eqnarray}
|\Psi\rangle=|vac\rangle+\int{d\Omega}F(\Omega)(a^{\dagger}_{H1}(\omega_{0}+\Omega)
a^{\dagger}_{V2}(\omega_{0}-\Omega){e^{i\Omega\tau_{0}}}\nonumber\\
+a^{\dagger}_{V1}(\omega_{0}+\Omega)a^{\dagger}_{H2}
(\omega_{0}-\Omega){e^{-i\Omega\tau_{0}}})|vac\rangle, \label{1}
\end{eqnarray}
where  $F(\Omega)$ is the spectral two-photon amplitude,  $\Omega$
is the frequency shift,  $a^{\dagger}_{\sigma i}$ are photon
creation operators in the horizontal and vertical polarization
modes (denoted by $\sigma=H,V$ ) and two spatial modes (denoted by
$i=1,2$); $\omega_0=\omega_p/2$, $\omega_p$ being the pump
frequency, and the terms corresponding to the case where both
photons of the two-photon state go to the same port of the
beamsplitter have been omitted. The difference between the group
velocities of signal and idler photons in the non-linear crystal
where SPDC occurs leads to the factors $e^{\pm i \Omega\tau_0}$ by
the two terms of Eq. (1), where $\tau_{0}=DL/2$, with $L$ being
the crystal length and $D\equiv\frac{1}{u_{V}}-\frac{1}{u_{H}}$
the inverse group velocity difference. To observe two-photon
interference in type-II SPDC, this delay $\tau_{0}$  has to be
compensated by means of a birefringent crystal. However, as one
will see from what follows, compensation is not necessary in the
presence of the fibre.

The coincidence counting rate can be calculated as

\begin{equation}
R_{c}\sim\int{dt}\int{d\tau}G^{(2)}(t,\tau),\label{2}
\end{equation}
where $G^{(2)}(t,\tau)$ is the second-order Glauber's correlation function defined as
\begin{equation}
G^{(2)}(t,\tau)\equiv\langle\Psi|E^{(-)}_{1}(t)E^{(-)}_{2}(t+\tau)
E^{(+)}_{2}(t+\tau)E^{(+)}_{1}(t)|\Psi\rangle,\label{2'}
\end{equation}
$E^{(\pm)}_{1,2}(t)$ are positive- and negative-frequency field operators at the
photodetectors, integration in $t$ is performed over the
measurement time interval, and integration in $\tau$, over the
coincidence circuit resolution window ~\cite{Mandel}. If SPDC is
obtained from a cw pump (the stationary case), $G^{(2)}$ does not depend on
$t$.

One can show ~\cite{Shih} that two-photon interference manifests
itself most explicitly in the difference between coincidence
counting rates for two cases. In the first case, both polarizers
are oriented at 45 degrees to the horizontal direction. In the
second case, polarizer 1 is oriented at $45^{\circ}$ and polarizer
2, at $-45^{\circ}$ to the same direction.

For these two configurations of the polarizers
($45^{\circ},45^{\circ}$ and $45^{\circ},-45^{\circ}$), the
correlation function takes the forms

\begin{equation}
G_{\pm}^{(2)}(\tau)=
|F(\tau-\tau_{0})\pm{F(\tau+\tau_{0})|^2}, \label{3}
\end{equation}
where $F(\tau)$  is the Fourier transform of the spectral
amplitude $F(\Omega)$  .

Note that for type-II SPDC the amplitude  $F(\tau)$ has a
rectangular shape with width  $2\tau_0$ ~\cite{Rubin}. Hence, if
the delay $\tau_0$ is not compensated, the two amplitudes in Eq.
(3) do not overlap and the interference is absent: the coincidence
counting rate, given by the integral of Eq.~(\ref{3}) over $\tau$,
 is the same for $(45^\circ, 45^\circ)$ and
$(45^\circ, -45^\circ)$ positions of the polarizers.

If the biphoton beam, before entering the beamsplitter, passes
through a sufficiently long fibre (with length $z$)
~\cite{sufflength}, the biphoton amplitude becomes
~\cite{Chekhova}

\begin{equation}
\widetilde{F}(\tau)=\frac{1}{\sqrt{4\pi i k''
z}}e^\frac{i(\tau-k'z)^2}{4k''z}F(\Omega)|_{\Omega=\frac{\tau-k'z}{2k''z}},
 \label{4}
\end{equation}
where $k',k''$ are, respectively, the first and the second
derivatives of the fibre dispersion law $k(\omega)$  at the
frequency $\omega_0$. Here, we assume that the group velocity
dispersion is the same for both polarizations.

Since for type-II SPDC  $F(\Omega)=\hbox{sinc}(\tau_0\Omega)$
~\cite{Shih}-~\cite{Pittman},
\begin{eqnarray}
G_{\pm}^{(2)}(\theta)\sim|e^{\frac{i(\theta-\tau_0)^2}{2\tau_0\tau_f}}\hbox{sinc}(\frac{\theta-\tau_0}{\tau_f})
\nonumber\\
\pm
e^{\frac{i(\theta+\tau_0)^2}{2\tau_0\tau_f}}\hbox{sinc}(\frac{\theta+\tau_0}{\tau_f})|^2,
 \label{5}
\end{eqnarray}
where $\theta\equiv\tau-k'z$ is the shifted time and
$\tau_f\equiv2k''z/\tau_0$  is the typical width of the
correlation function after the fibre ~\cite{Krivitsky}.

Taking into account that $\tau_f>>\tau_0$ , one can rewrite Eq.
(6) as
\begin{eqnarray}
G_{-}^{(2)}(\theta)\sim\frac{\sin^4(\theta/\tau_f)}{(\theta/\tau_f)^2},
\nonumber\\
G_{+}^{(2)}(\theta)\sim\frac{\sin^2(\theta/\tau_f)\cos^2(\theta/\tau_f)}{(\theta/\tau_f)^2}.
\label{6}
\end{eqnarray}

Thus, in the presence of the fibre, two-photon interference can
manifest itself in the shape of the second-order Glauber's
correlation function. However, if the correlation function is
integrated over all delays $\theta$ (which is the case in most
experiments on two-photon coincidence counting), the result will
be the same for both configurations of the polarizers. To see the
structure given by Eq.s~(\ref{6}), one has to measure the
correlation function with a sufficiently good time resolution. Two
main reasons for the resolution reduction must be considered: the
time jitter of the detectors (typically of the order of several
hundreds of picoseconds), which sets the ultimate time resolution
of the measurements, and the jitter contribution of the
electronics (amplitude walk and noise) used for the coincidence
detection.

Coincidence detection by means of a coincidence logic gate circuit
typically shows resolving time, for commercial available devices,
of a few nanoseconds, which can be much greater than the spread
$\tau_f$ of the correlation function because of the propagation in
a dispersive fibre.

Another widely used coincidence detection technique, which will be
further called the START-STOP method, involves direct measurement
of the time delay between the photocount pulses of the two
detectors by means of a time-to-amplitude converter (TAC), which
converts linearly the time interval between the two input pulses
(START and STOP) into an output pulse of a proportional amplitude.
This analog pulse is forwarded to a multichannel analyzer (MCA),
which gives the histogram of the input pulse amplitudes
corresponding to the probability distribution for the time
interval between the counts of the two detectors. One can show
~\cite{Mandel} that in the limit of small photon fluxes, this time
interval distribution coincides in shape with the second-order
intensity correlation function. The resolution of such technique
could be on the order of one picosecond if the detector jitter
time were negligible.
The expected experimental coincidence distribution, in the case of
negligible jitter time contribution from the experimental
apparatus, was calculated numerically (Fig.2a) for the following
parameters: $k''=3.2\cdot10^{-28}\hbox{s}^2\hbox{/cm}$, fibre
length $z=250\hbox{m}$, $D=1.5\hbox{ps/cm}$, crystal length
$L=0.05 \hbox{cm}$. The corresponding FWHM of the correlation
function after the fibre is $1.2$ ns ($\tau_f=0.43\hbox{ns}$).

One can see that for the polarizers set at ($45^\circ,-45^\circ$),
destructive interference occurs at the center of the peak. On the
other hand, for the polarizers set at ($45^\circ,45^\circ$), there
is constructive interference at the centre of the peak.
High-visibility polarization interference can be observed in this
case by selecting only several channels of the MCA corresponding
to the center of the peak and rotating one of the polarizers.

If we take into account the time jitter of the experimental setup,
the interference structure gets smeared.  In Fig.2b we show the
numerical evaluation of the coincidence distribution with an
account for the time jitter of the experimental apparatus, which
is the convolution of the peak plotted in Fig.2a with a 750 ps (a
typical jitter for two commercial APD detectors) FWHM Gaussian
function. One can see that in the case of Fig.2b, the change in
the number of coincidences observed at zero delay for
($45^\circ$,$45^\circ$) and ($45^\circ$, $-45^\circ$) orientations
of the polarisers corresponds to only 45\% visibility, unlike in
the case of Fig.2a, where almost 100\% visibility is observed.

It is worth hinting to what happens if the e-o delay $\tau_0$ is
compensated partly, by means of a birefringent material
introducing an opposite-sign delay. Here \cite{forth} we only
mention that the period of the sine and cosine functions in
Eqs.~(\ref{6})  increases above the width of the envelope and thus
the integral numbers of coincidences corresponding to the two
polarizer configurations  differ. As a result, there is nonzero
visibility even in the absence of the correlation function time
selection. For complete compensation, the visibility would be
$100$\%.

In order to experimentally observe two-photon interference in the
shape of the correlation function of biphotons, we have generated
biphoton pairs via SPDC by pumping a type-II 0.5 mm BBO crystal
with a 0.1 Watt CW $\hbox{Ar}^{+}$ laser beam at the wavelength
351 nm in the collinear frequency-degenerate regime. No
compensating birefringent material was used after the crystal.
Then the pump beam was eliminated by a high-reflectivity  UV
mirror plus an anti-UV cutoff filter and the SPDC radiation was
coupled to a 250 m - long single-mode fibre with FMD$=4\mu$ by a
20x microscope objective lens. In order to increase the efficiency
of coupling SPDC radiation into the fibre, the pump was focused
into the crystal using a lens with the focal length 13.5 cm.
Because of the rapid polarization drift in the fibre, a fibre
polarization controller was introduced at the output. After the
fibre, the biphoton pairs were addressed to a nonpolarizing
beamsplitter and two photodetection apparatuses (consisting of
polarisers, red-glass filters and APDs). The photocount pulses of
the two detectors, after passing through delay lines, were
analyzed by means of the START-STOP method, and the second-order
correlation function, $G^{(2)}(\theta)$, was observed at the MCA
output. The FWHM of the coincidence peak in the absence of the
fibre was measured to be 0.75 ns, a value substantially determined
by the APD time jitter. The resolution determined by the MCA
channel width, in our case 2.5 ps, is much smaller and therefore
negligible.

If the biphotons propagate through the fibre, the FWHM grows up to
$1.2$ ns. The peak observed without polarization selection is
shown in Fig.3a.  In the presence of the polarizers, the shape of
the peak changes (Fig.3b,c). For the ($45^\circ$;$-45^\circ$)
settings of the polarizers, destructive interference in the middle
of the peak leads to a shape in agreement with the calculated one
(Fig.2b). For the polarizers set at ($45^\circ$;$45^\circ$), as
expected, there is a maximum at the center of the peak. If only
the central part of the peak is selected, the observed
 variation in the number of coincidences due to the change in the polarizer settings
 corresponds to 35\%
 visibility of polarization interference.

In summary, we have observed effects of two-photon interference in
the shape of the second-order Glauber's correlation function for
biphoton light generated via type-II SPDC and transmitted through
an optical fibre.
It is worth noticing that our results show how, for a sufficiently
long fibre (hundreds of meters), a Bell state can be generated
without compensating for the e-o delay between the signal and the
idler photons. In this particular method of Bell states
preparation, distinguishability of the interfering two-photon
amplitudes is erased by spreading them in time and by selecting
coincidences within a suitable time window. Even if this method
could appear more complicated than the traditional one, based on
using additional birefringent elements, it is interesting as a
conceptually different way for restoring entanglement. In this
sense it must be noticed that selection of the coincidence time
window is similar to the frequency selection, which can be
achieved by using narrow-band filters in some experiments on
Bell-state preparation ~\cite{Gisin,Shih2}, since the
group-velocity dispersion of the fibre actually performs the
Fourier transformation of the biphoton spectrum into the
correlation function ~\cite{Chekhova}. However, this does not mean
that the observed interference is present in the spectrum of the
down-converted photons, since the observed shape of the
correlation function substantially depends on the polarizer
settings in both arms, which, in the general case, are different.
Another distinction \cite{forth} from the experiments where
narrow-band filters are used for the Bell states preparation
consists of the fact that, in our experiment, one can achieve high
visibility by selecting not necessarily the central part of the
two-photon peak but any symmetric side parts where the coincidence
counting rates differ for different polarizer settings (see
Fig.2). Furthermore, one should notice that these effects
inevitably occur in any quantum communication experiment where
biphoton light is transmitted through optical fibres, and if the
fibres are sufficiently long and temporal selection of the
coincidences is provided, there is no need for o-e delay
compensation.

Finally, we would like to mention that the measurement of the
observed broadening of the second-order Glauber correlation
function could eventually be used for the evaluation of the
optical fibre chromatic dispersion (in analogy to Ref.
~\cite{Brendel}).

\section{Acknowledgments}
This work was supported by MIUR (FIRB RBAU01L5AZ-002), by "Regione
Piemonte E14", by Fondazione San Paolo, and by the Russian
Foundation for Basic Research, grant No. 05-02-16391. One of us
(L.K.) acknowledges the support of Intas Grant for young
scientists No. 03-55-1971.

\begin{figure}
\includegraphics[height=3cm]{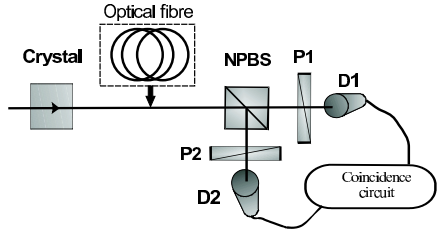} \caption{A setup for
observing two-photon polarization interference for type-II SPDC.
The fibre is introduced for spreading the correlation function and
observing its interference structure.}

\end{figure}

\begin{figure}
\includegraphics[height=4cm]{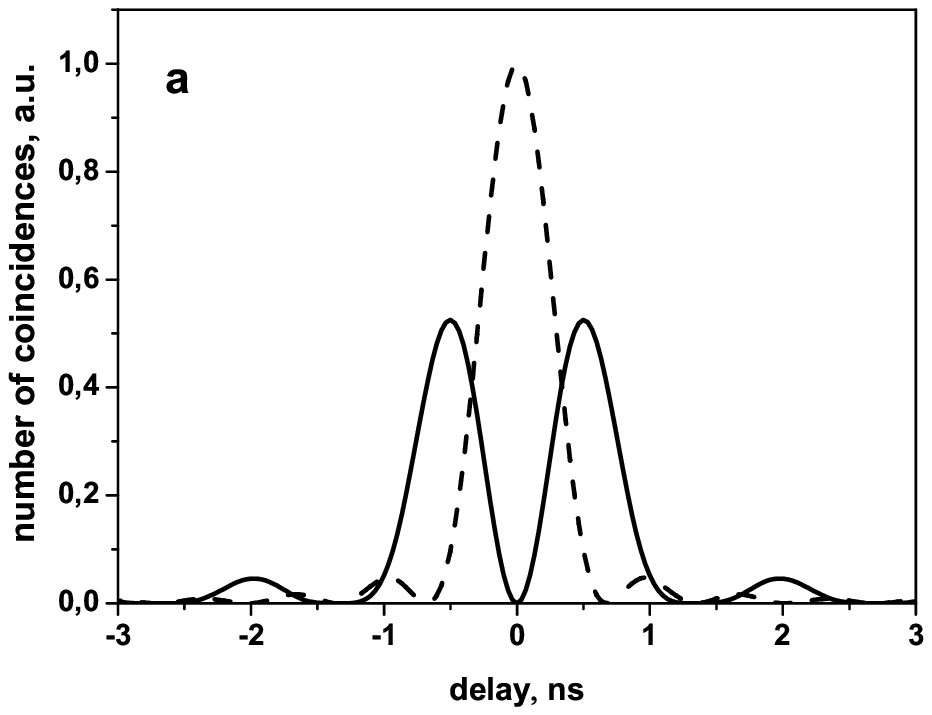}
\includegraphics[height=4cm]{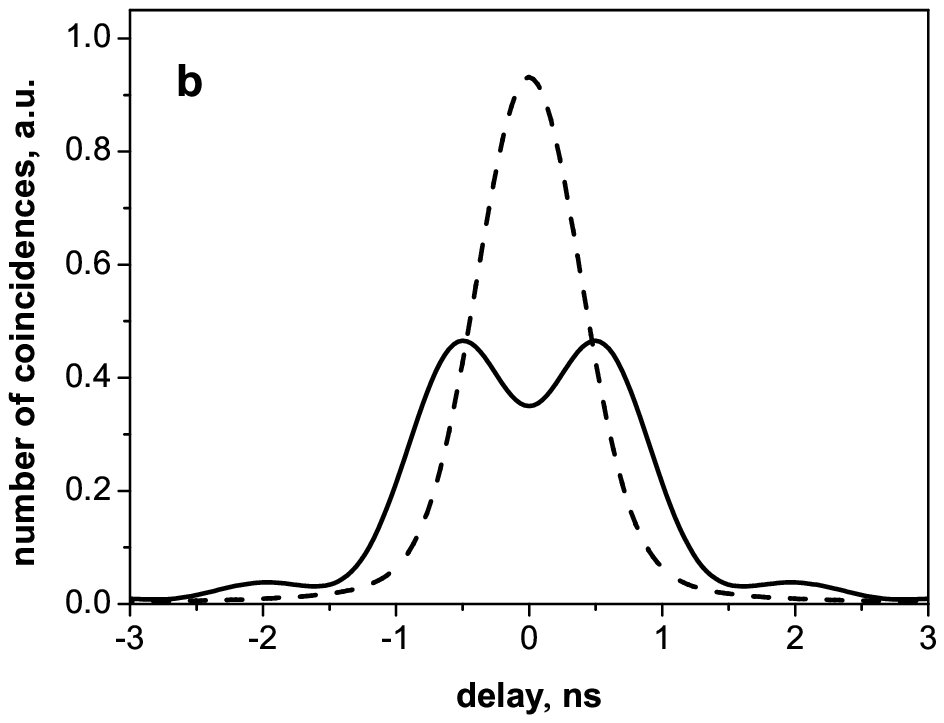} \caption{Time delay distribution of coincidences
calculated for $k''=3.2\cdot10^{-28}\hbox{s}^2/\hbox{cm}$,
$z=250\hbox{m}$, $D=1.5 \hbox{ps/cm}$, $L=0.05\hbox{cm}$, and the
polarizers oriented at ($45^\circ, -45^\circ$) (solid line) and at
($45^\circ,45^\circ$) (dashed line) (a) for an ideal measurement
setup with negligible time jitter; (b) for a measurement setup with
a time jitter of 750 ps. As the mutual orientation of the polarizers
changes from parallel to orthogonal, the change in the coincidence
number at zero delay corresponds to 100\% visibility in the (a) case
and to 45\% visibility in the (b) case.}
\end{figure}



\begin{figure}
\includegraphics[height=4cm]{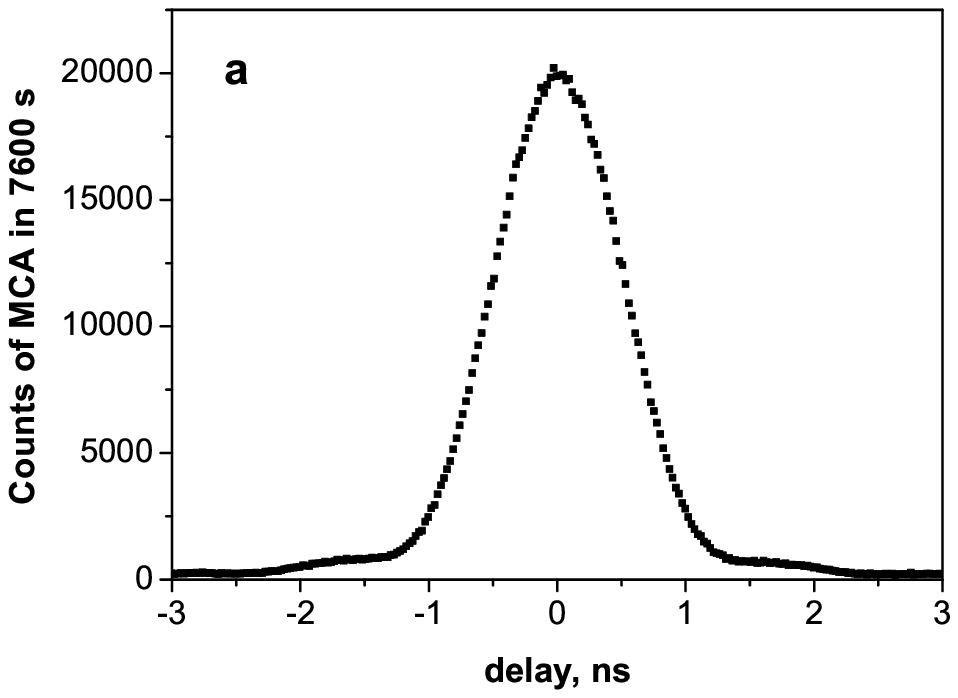}
\includegraphics[height=4cm]{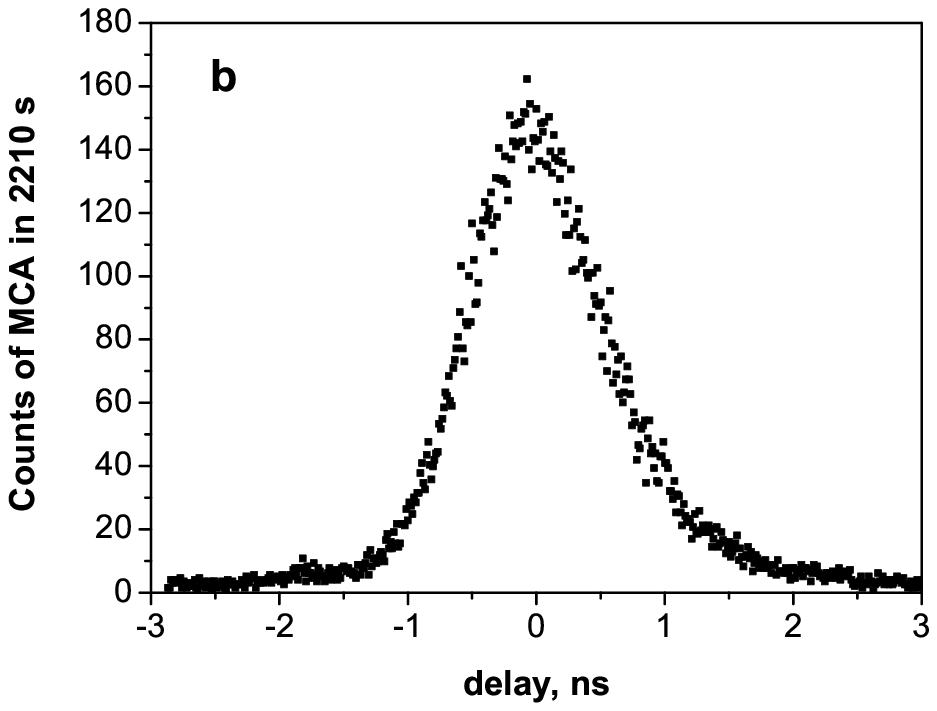}
\includegraphics[height=4cm]{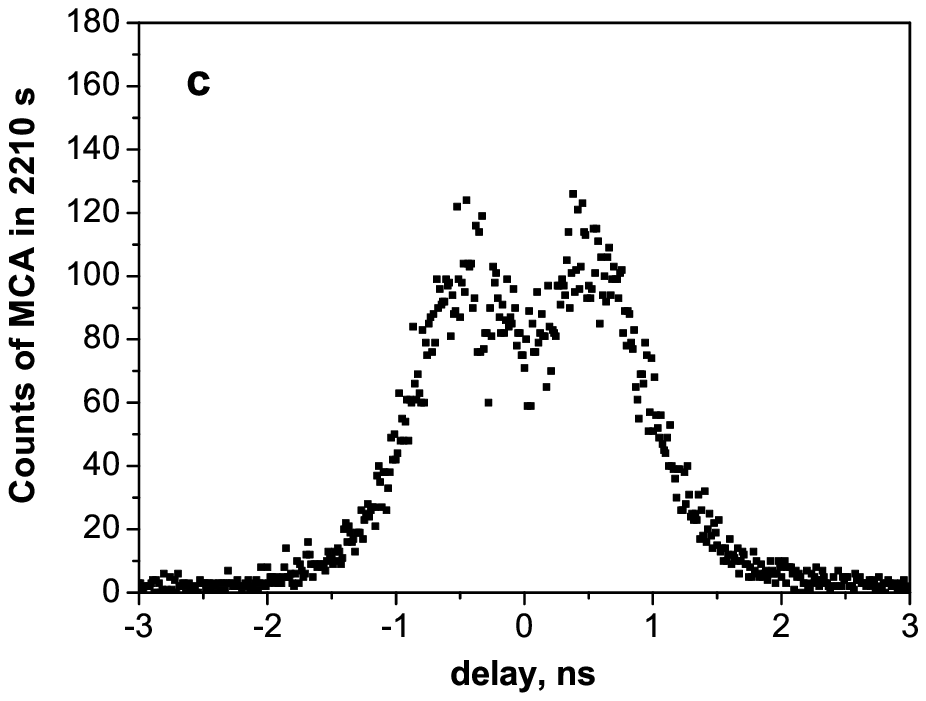}
 \caption{  Observed time delay distribution of coincidences:(a)without the
 polarizers;(b) with polarizers set at ($45^{\circ},45^{\circ}$), which corresponds to measuring
$G_{+}^{(2)}(\theta)$; (c) with polarizers set at
($45^{\circ},-45^{\circ}$), which corresponds to measuring
$G_{-}^{(2)}(\theta)$. Destructive interference at the center of
the coincidence peak is clearly seen. As the parallel polarizer
configuration (b) is switched to the orthogonal polarizer
configuration (c), the change in the coincidence number at the
center of the peak corresponds to approximately 35\% visibility.}
\end{figure}


\begin{references}
\bibitem{sufflength} A typical length of about 1
meter is sufficient.
\end{references}
\end{document}